\documentclass[aps,pra,reprint]{revtex4-1}

\usepackage{amsmath,amsfonts,amssymb,graphicx,braket,bbold,mathtools,array}
\usepackage{multirow}
\usepackage[]{units}
\usepackage{hyperref}
\hypersetup{
    colorlinks=true,
    linkcolor=blue,
    filecolor=blue,      
    urlcolor=blue,
    citecolor=blue
}

\usepackage[table]{xcolor}

\usepackage[T1]{fontenc}
\usepackage{tikz}
\usetikzlibrary{calc,patterns,decorations.pathmorphing,decorations.markings, shapes, positioning, shapes.geometric}

\makeatletter
      \def\CT@@do@color{%
      \global\let\CT@do@color\relax
            \@tempdima\wd\z@
            \advance\@tempdima\@tempdimb
            \advance\@tempdima\@tempdimc
    \advance\@tempdimb\tabcolsep
    \advance\@tempdimc\tabcolsep
    \advance\@tempdima2\tabcolsep
            \kern-\@tempdimb
            \leaders\vrule
                    \hskip\@tempdima\@plus  1fill
            \kern-\@tempdimc
            \hskip-\wd\z@ \@plus -1fill }
    \makeatother

\newcommand*{\figref}[2][]{%
  \hyperref[{fig:#2}]{%
   \ref*{fig:#2}%
    \ifx\\#1\\%
    \else #1%
    \fi
  }%
}

\begin{document}
\title{g-tensor resonance in double quantum dots with site-dependent g-tensors}
\author{Philipp Mutter}
\email{philipp.mutter@uni-konstanz.de}
\author{Guido Burkard}
\email{guido.burkard@uni-konstanz.de}
\affiliation{Department of Physics, University of Konstanz, D-78457 Konstanz, Germany}

\begin{abstract}
Pauli spin blockade (PSB) has long been an important tool for spin read-out in double quantum dot (DQD) systems with interdot tunneling $t$. In this paper we show that the blockade is lifted if the two dots experience distinct effective magnetic fields caused by site-dependent g-tensors $g_L$ and $g_R$ for the left and right dot, and that this effect can be more pronounced than the leakage current due to the spin-orbit interaction (SOI) via spin-flip tunneling and the hyperfine interaction (HFI) of the electron spin with the host nuclear spins. 
Using analytical results obtained in special parameter regimes, we show that information about both the out-of-plane and in-plane g-factors of the dots can be inferred from characteristic features of the magneto-transport curve. 
For a symmetric DQD, we predict a pronounced maximum in the leakage current at the characteristic out-of-plane magnetic field $B^* = t/ \mu_B \sqrt{g_z^L g_z^R}$ which we term the g-tensor resonance of the system.
Moreover, we extend the results to contain the effects of strong SOI and argue that in this more general case the leakage current carries information about the g-tensor components and SOI of the system.
\end{abstract}

\maketitle
\section{Introduction}\label{sec:Introduction}
The basic unit in quantum information processing is the qubit, the quantum version of a classical bit. A prominent physical realization of a qubit is the spin of an electron \cite{LossandDiVincenzo} as it provides a natural two-state system.  A key requirement \cite{DiVincenzo2000} of a quantum computer built from such spin qubits is the ability to read out the spin states. PSB has proven to be an important tool for this purpose \cite{Petta2005, Koppens2005, Koppens2006, Hanson2007, Nowack2007}. Consider a semiconductor DQD tuned to the (0,1) - (1,1) - (0,2) triple point in the charge stability diagram, where ($n$,$m$) denotes $n$ electrons in the left and $m$ electrons in the right dot. Such configurations are regularly achieved in experiment \cite{Katsaros2011, Watzinger2018}. In a (1,1) $\rightarrow$ (0,2) $\rightarrow$ (0,1) $\rightarrow$ (1,1) transport cycle PSB in DQD systems occurs when a triplet state is formed in the (1,1) charge configuration. The (0,2) triplet state typically has a higher energy than the singlet and is out of reach in the PSB regime, while the transition to the (0,2) singlet state is forbidden by spin conservation. This situation, where current through the DQD system is prevented from flowing is known as (Pauli) spin blockade. A rotation of one of the electron spins then leads to a measurable electric current, or for a closed DQD system to the motion of a single electron which can be read out via charge sensing, i.e., the PSB amounts to a spin-to-charge conversion scheme. However, the blockade can be lifted by various mechanisms influencing electron spins. The resulting leakage current due to the HFI \cite{Jouravlev2006, Palyi2009}, the SOI \cite{Danon2009, Palyi2010} and disorder in spin-valley systems \cite{Palyi2010} has been derived theoretically and observed in experiment \cite{Buitelaar2008, Churchill2009a, Churchill2009b,Nadj-Perge2010, Li2015}. In this paper, we investigate spin blockade lifting as a result of site-dependent g-tensors. Such a situation is observed, e.g., for holes in Ge \cite{hendrickx2019fast, Hendrickx2020} where the g-tensors are also highly anisotropic due to the strong SOI \cite{Winkler2003}. In earlier works on spin blockade lifting in a DQD with SOI \cite{Danon2009} the g-tensor renormalization was taken to be equal in both dots. Since in this case the leakage current can be regarded as a function of the Zeeman energy, $E_Z =\mu_B | g \mathbf{B}|$, with common g-tensor $g$ and magnetic field $B$, the g-tensors do not qualitatively change the magneto-transport curve. When the g-tensors in the two dots differ, however, we expect a qualitative change of the leakage current. Indeed, different g-tensors cause site-dependent effective magnetic fields acting on the two dots and thereby allow for an interaction Hamiltonian that does not conserve the total spin. Consequently, previously blocked transitions between the triplet  and singlet states are enabled leading to a lifting of the PSB. Note that different g-tensors are caused by different constitutions of the two dots in a DQD due to site-dependent confinement, strain or material composition \cite{Kiselev1998, Nenashev2003, Nakaoka2004, Nakaoka2005, Oliveira2008, Aleshkin2008}. While these differences are expected to occur in any realistic DQD system due to unavoidable imperfections in quantum dot growth or engineering, we expect the difference in the g tensors to be particularly noticeable in materials with a strong SOI. Hence, spin blockade lifting due to different g-tensors is to be distinguished from pure spin-orbit induced spin blockade lifting as the latter persists even for identical dots while the former does not.

In the process of investigating the lifting of the PSB, we will find that the associated magneto-transport signal conveys information about the g-tensors of the charge carriers in the dots. The nature of the g-tensors is of interest since they play a prominent role in qubit manipulations, another key requirement of a quantum computer \cite{DiVincenzo2000}. Tunable g-factors have been observed \cite{Salis2001, Kato2003, Jovanov2011, Deacon2011, Ares2012, Maier2013}, and it is therefore possible that g-factor control over single spins can be achieved.  Recently it was reported that knowledge of the g-tensors in a DQD is also important in building and manipulating singlet-triplet qubits \cite{hofmann2019assessing}. As it touches on the topics of both spin manipulation and read-out, we hope that the investigation of the effects of site-dependent g-tensors on the magneto-transport curve and the information it contains about the g-tensors in the dots provides new insights into quantum dot materials and thus facilitates fabrication and manipulation of quantum information devices.

The remainder of this work is structured as follows: In Sec.~\ref{sec:modelandmethods} we introduce the basic model and tools needed to obtain the leakage current. In Sec.~\ref{sec:Idifferentg} we present numerical and analytical results for the form of the leakage current. The magnitude of the magnetic field when the current is maximal will supply us with a way of accessing information about the g-tensors in the dots experimentally. These results are generalized to systems with strong SOI in Sec.~\ref{sec:strongSOI}. Finally, Sec.~\ref{sec:conclusion} summarizes the results and provides a conclusion.

\section{Model and methods}\label{sec:modelandmethods}
\begin{figure}
		\includegraphics[width=0.9\linewidth]{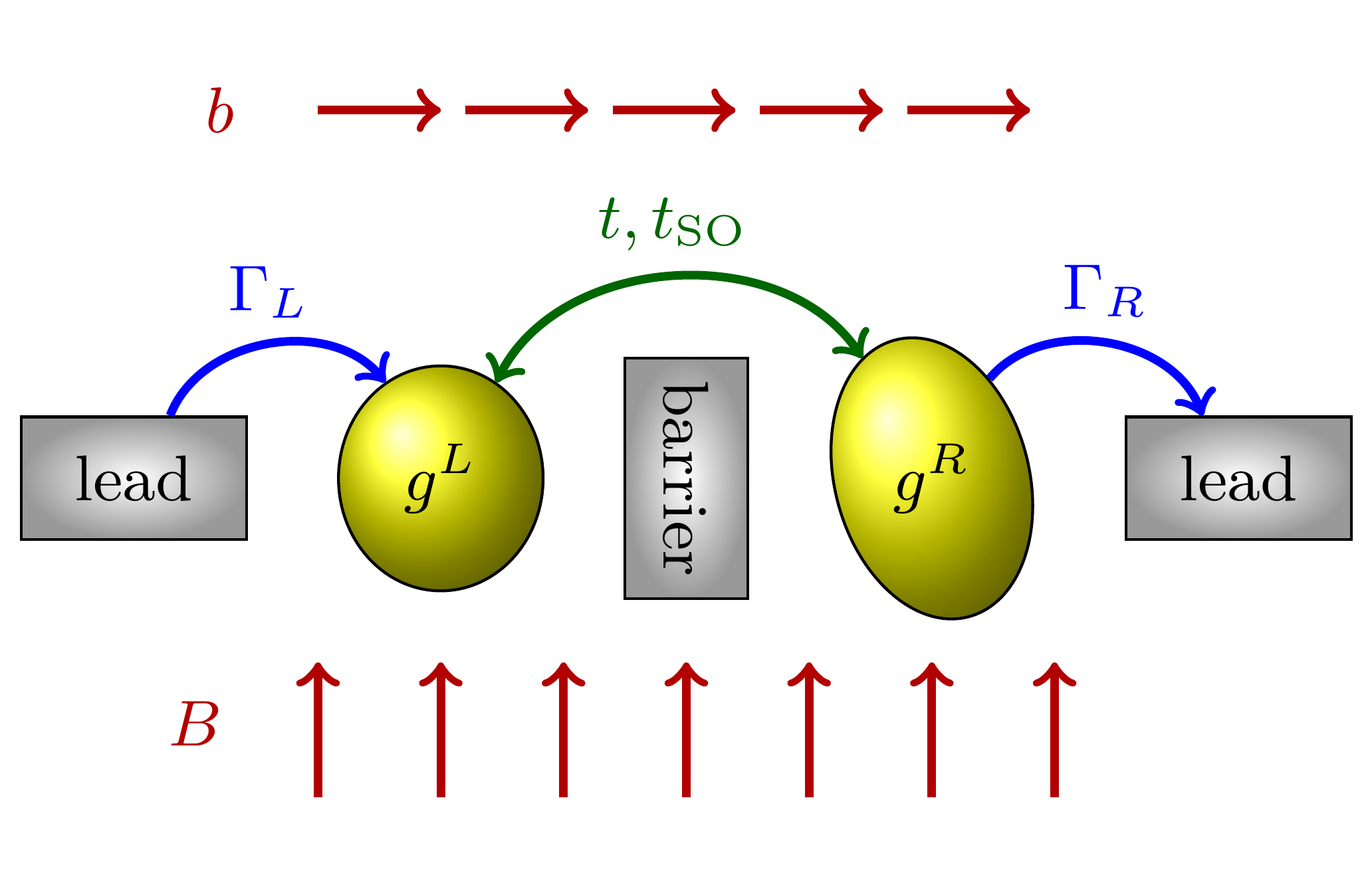}
	    \caption{The basic DQD setup for PSB. In the presence of both in-plane and out-of-plane magnetic fields we observe a lifting of the spin blockade since site-dependent g-tensors cause different effective magnetic fields in the two dots. The (anisotropic) g-tensors in the dots are depicted by ellipsoids with the components of the diagonal g-tensors as the principal axes. While we assume equal principal systems in this paper, the g-tensors may in general be diagonal in different bases.}
	    \label{fig:transportsetup}
	\end{figure}

We consider a transport setup as shown in Fig.~\ref{fig:transportsetup}, consisting of a DQD connected to leads. When a bias voltage is applied, charge carriers may tunnel from the left lead to the left dot with a rate $\Gamma_L$. The left and right dots are separated by an energy barrier through which charge carriers may tunnel with a tunneling matrix element $t$. Later, we will also allow for spin-orbit induced tunneling with a tunneling matrix element $t_{\text{SO}}$ and spin relaxation within the (1,1) charge configuration with a rate $\Gamma_{\text{rel}}$. Once in the right dot, charge carriers can tunnel to the right lead with a rate $\Gamma_R$, thereby leaving the DQD system. 
Introducing the detuning $\epsilon$ between the (0,2) and (1,1) charge states and setting the energy of the latter to zero, the basic Hamiltonian describing the DQD reads
	\begin{equation}
	\label{eq:HDQD}
		H_{\text{DQD}} = \epsilon \vert S_{02}\rangle \langle S_{02} \vert + t\left( \vert S\rangle \langle S_{02} \vert +  \vert S_{02}\rangle \langle S \vert \right).
	\end{equation}
In Eq.~\eqref{eq:HDQD} we have defined the two-particle singlet states
	\begin{align*}
		&\vert S_{02} \rangle = c_{R \uparrow}^{\dagger} c_{R \downarrow}^{\dagger}  \vert \text{vac} \rangle, \\
		&\vert S \rangle = \frac{1}{\sqrt{2}} \left(c_{L \uparrow }^{\dagger} c_{R \downarrow}^{\dagger} -  c_{L \downarrow }^{\dagger} c_{R \uparrow}^{\dagger}  \right) \vert \text{vac} \rangle,
	\end{align*}
where the operator $c_{d \sigma}^{\dagger}$ creates a particle of (pseudo-)spin ${\sigma\in\lbrace\uparrow,\downarrow\rbrace}$ in the left ($d = L$) or right ($d = R$) dot, and $\vert \text{vac} \rangle$ is the vacuum state.  For heavy holes in cubic semiconductors such as Ge, the pseudo-spin description refers to the states with total angular momentum projection $j_z=\pm 3/2$. Note that we may neglect the singlet state in the (2,0) charge configuration for the same reason that allowed us to disregard the (0,2) triplet state, i.e., since the (2,0) singlet is energetically out-of-reach in the PSB regime. To describe the interaction of the electrons with the external field we propose a Zeeman Hamiltonian including  site-dependent g-tensors for a DQD in a magnetic field $\boldsymbol{B}$,
	\begin{equation}
	\label{eq:HZ}
		H_{\text{Z}} = \sum_{d}   \sum_{i,j} B_i  g_{ij}^d S^d_j  
		 = \sum_{d} \boldsymbol{\mathcal{B}}^d  \cdot \boldsymbol{S}^d .
	\end{equation}
The expression features the effective magnetic field in dot $d$, $\mathcal{B}_i^d \equiv \sum_j g_{ij}^d B_j$,  and the spin operator in dot $d$, $\boldsymbol{S}^d = \boldsymbol{ \sigma}^d /2$, where $\boldsymbol {\sigma} = (\sigma_x, \sigma_y, \sigma_z)^T$ is a vector containing the three Pauli matrices. The indices $i$ and $j$ run over the three spatial directions, and the index $d$ runs over the left ($L$) and right ($R$) dots. We have set $\mu_B = \hbar = 1$, thus measuring rates and magnetic fields in units of energy. The form of the Hamiltonian in Eq.~\eqref{eq:HZ} is similar to that of the Hamiltonian proposed in Ref.~\cite{Jouravlev2006}. The physics it describes, however, is rather different: In the cited paper the external magnetic field enters as a means to detect the signature of the leakage current caused by the intrinsic HFI. Here, the source of the leakage current \textit{is} the externally applied magnetic field acting differently on the two dots due to site-dependent g-tensors. This difference is obvious when considering the zero-field values of the electric current: In the absence of external fields the (averaged) current in the referenced work is finite, while it is zero in the present investigation.

Even though we show the leakage current to be the result of different g-tensors alone, we may include the HFI as it is expected to be present when dealing with materials containing at least one nuclear species with a finite nuclear spin (GaAs, as well as Si and Ge without isotopic purification). Therefore, we add a term describing the HFI of the host nuclear spins with the electron spin in the DQD to the Hamiltonian,
	\begin{align}
	\label{eq:Hhfi}
		H_{\text{HFI}} = \sum_{d,i,j}  \sum_{n=1}^{N_d}  I^d_{n,i} A_{n,ij}^d S_j^d,
	\end{align}
where $I^d_{n,i}$ denotes the $i^{\text{th}}$ component of the nuclear spin operator of the $n^{\text{th}}$ of $N_d$ nuclei in dot $d \in \lbrace L,R \rbrace$. Note that the hyperfine tensor $A_{n,ij}^d$ is not necessarily isotropic \cite{Witzel2007, Hodges2008, Zhang2011}, e.g., for holes it can be highly anisotropic due to the p-symmetry of the heavy and light hole wave functions in the valence band \cite{Testelin2009, Greilich2011, Carter2014}. The nuclear fields are assumed to be normally distributed with zero mean \cite{Khaetskii2002, Merkulov2002, Coish2004}. Later on we will look for regimes in which the HFI can be neglected. To this end, the variance of the $i^{\text{th}}$ component of the effective nuclear magnetic field in dot $d$, $\mathcal{K}_i^d$, may be estimated via $\langle ( \mathcal{K}_i^d)^2 \rangle = \sum_{n,j} \langle A_{n,ij}^d  I_{n,j}^d \rangle^2/N_{\text{eff}}$ for $i,j \in \lbrace x,y,z \rbrace$ in accordance with the central limit theorem \cite{Merkulov2002, burkardreview2}. Here, $N_{\text{eff}}$ denotes the effective number of nuclei in the quantum dot. The nuclear fields are assumed to be independent and quasistatic, i.e., static with respect to the much shorter DQD time scales. The time scale on which the nuclear spins change, however, is expected to be much smaller than the typical time of a measurement, and we will thus average the current over many realizations of the nuclear fields later on.

Finally, we allow for non spin-conserving tunnel couplings caused by the SOI. Defining the two-particle triplet states,
	\begin{align*}
		&\vert T_0 \rangle = \frac{1}{\sqrt{2}} \left(c_{L \uparrow }^{\dagger} c_{R \downarrow}^{\dagger} +  c_{L \downarrow }^{\dagger} c_{R \uparrow}^{\dagger}  \right) \vert \text{vac} \rangle, \\
		&\vert T_+ \rangle = c_{L \uparrow}^{\dagger} c_{R \uparrow}^{\dagger}  \vert \text{vac} \rangle, \\
		&\vert T_- \rangle = c_{L \downarrow}^{\dagger} c_{R \downarrow}^{\dagger}  \vert \text{vac} \rangle,
	\end{align*}
the part of the Hamiltonian describing the SOI reads 
	\begin{align}
	\label{eq:HSO}
		H_{\text{SOI}} = i t_z \vert T_0 \rangle \langle S_{02} \vert + \sum_{\pm} \mp i t_{\mp} \vert T_\pm \rangle \langle S_{02} \vert  + \text{h.c.},
	\end{align}
where $t_{\pm} = (t_x \pm i t_y)/ \sqrt{2}$ \cite{Danon2009}. The vector containing the spin-orbit induced tunnel matrix elements $\boldsymbol{t}_{SO} = (t_x, t_y, t_z)^T$ transforms as a real vector under coordinate transformations, and we refer to it as the spin-orbit vector of the system. The magnitude can be approximated as  $\vert \boldsymbol{t}_{\text{SO}} \vert / t \equiv t_{\text{SO}}/t \approx r_{\text{dot}}/l_{\text{SO}}$, where $ r_{\text{dot}}$ is the typical dot radius and $ l_{\text{SO}}$ is the particle's spin-orbit length \cite{Nadj-Perge2010}. Note that the effective couplings described by $\boldsymbol{t}_{\text{SO}}$ are obtained by considering spin-orbit induced transitions from the triplets $\vert T_{0, \pm} \rangle$ to the singlet $\vert S \rangle$ in the (1,1) configuration and subsequent transitions $\vert S \rangle \rightarrow \vert S_{02} \rangle$ with tunneling matrix element $t$. Therefore, we expect $t_{SO} < t$.

We use a master equation to calculate the leakage current numerically \cite{Jouravlev2006}. To describe PSB, we assume throughout this work that the right dot-lead tunneling rate $\Gamma_R$ is much smaller than the left lead-dot tunneling rate $\Gamma_L$, such that the current is determined by the (0,2) charge configuration. Moreover, we assume the inter-dot tunneling to occur faster than the tunneling to the right lead, $\Gamma_R < t$. This `bottleneck' setup is common in PSB transport investigations \cite{Palyi2010}.  Denoting the full Hamiltonian by $H = H_{\text{DQD}} + H_{\text{Z}} + H_{\text{HFI}} + H_{\text{SOI}}$, we consider the following master equation in the Lindblad form for the density matrix $\rho$, 
	\begin{align}
	\label{mastereq}
		\frac{\partial \rho}{\partial t} = -i [H, \rho] + \boldsymbol{\Gamma}_R [\rho]  + \boldsymbol{\Gamma}_{\text{rel}}[\rho],
	\end{align}
where the dissipation operators describing the tunneling to and from the leads, $\boldsymbol{\Gamma}_R [\rho] $, and low-temperature relaxation processes, 
$\boldsymbol{\Gamma}_{\text{rel}}[\rho]$, are given by
	\begin{subequations}
	\begin{align}
	\label{eq:tunnelingtolead}
		& \boldsymbol{\Gamma}_R[\rho] = \frac{\Gamma_R}{4} \sum_i \left( L_i \rho L_i^{\dagger}  - \frac{1}{2} \left\lbrace L_i^{\dagger}  L_i,   \rho \right\rbrace \right), \\
	\label{eq:relaxation}
		& \boldsymbol{\Gamma_{\text{rel}}}[\rho] = \frac{\Gamma_{\text{rel}}}{2} \sum_d \left( S_-^d \rho S_+^d  - \frac{1}{2} \left\lbrace S_+^d S_-^d,   \rho \right\rbrace \right),
	\end{align}
	\end{subequations}
Here, $\lbrace a,b \rbrace = ab + ba $ denotes the anti-commutator, the four Lindblad (quantum jump) operators $L_i$ are given by $\vert S \rangle \langle S_{02} \vert$ and  $\vert T_{0, \pm} \rangle \langle S_{02} \vert$, and $S_{\pm}^d = S_x^d \pm iS_y^d$ are the spin raising and lowering operators in dot $d \in \lbrace L,R \rbrace$. In Eq.~\eqref{eq:tunnelingtolead} we assume that starting from the $\vert S_{02}\rangle$ state, the electron may tunnel to the right lead with a rate $\Gamma_R$ and that all states in the (1,1) charge configuration are subsequently refilled with a rate $\Gamma_R/4$. Motivated by recent experiments \cite{Watzinger2018, hofmann2019assessing, Hendrickx2020}, we assume thermal energies well below the Zeeman energy of the system and spin-flips in each dot with a spin relaxation rate $\Gamma_{\text{rel}}/2$ in Eq.~\eqref{eq:relaxation}. Relevant relaxation processes may include interactions with nuclear spins \cite{Erlingson2002}, spin-orbit-phonon induced spin flips \cite{Flindt2006}, spin-spin interactions \cite{Trif2008}, cotunneling or spin exchange with the leads \cite{Nadj-Perge2010}. The set of equations in Eq.~\eqref{mastereq} is closed by the normalization condition, $\text{Tr} \rho = 1$, and the current through the DQD is given by $I = e \Gamma_R \rho_{S_{02}S_{02}}$, where $\rho_{S_{02}S_{02}}$ is the diagonal component of the density matrix describing the probability for the system residing in the 
(0,2) charge configuration in the steady state ($\partial \rho / \partial t = 0$).

Seeking physical insights into the numerical results, we use a rate equation approach in order to analytically determine the leakage current. Following Ref.~\cite{Coish2011}, the current is given by
	\begin{equation}
	\label{eq:current}
		\frac{I}{e}= \left( \sum_i \frac{n_i}{\Gamma_i} \right)^{-1}.
	\end{equation}
Here, $n_i$ is the average number of times a state $i$ is occupied per transport cycle, and $\Gamma_i$ is the rate of this state transitioning into the (0,2) singlet.  The rates may be obtained in special cases up to first order perturbation theory as the transition rates of the occupied states into the (0,2) singlet,
	\begin{equation}
	\label{eq:rates}
		\Gamma_j = \Gamma_R \left\vert  \langle S_{02} \vert j \rangle +  \sum_{\alpha \neq j} \frac{\langle  S_{02} \vert \alpha\rangle \langle \alpha \vert H_1 \vert j \rangle }{E_j - E_{\alpha}}  \right\vert^2,
	\end{equation}
where the states $\vert j \rangle$, $\vert \alpha \rangle$ are eigenstates of the dominant Hamiltonian $H_0 = H - H_1$. The exact form of the perturbation $H_1$ has to be chosen according to the regime under consideration. In the following section we compare the numerical leakage current with analytical expressions obtained in special parameter cases. 
\section{Form of the leakage current}\label{sec:Idifferentg}
In this section we investigate the form of the leakage current caused by different g-tensors. When the two dots constituting the DQD are formed in the same material, it is reasonable to assume the two g-tensors to be diagonal in the same basis, for instance the cubic basis of the crystal for the case of holes in Ge \cite{Ares2012, Watzinger2016}. Inspired by recent experiments \cite{hofmann2019assessing}, we consider lateral quantum dots, in which the g-tensor in the in-plane subspace is degenerate, i.e., $g_x = g_y $. The coordinate system may be chosen such that the total external magnetic field assumes the form $\boldsymbol{B} = (b,0,B)^T$ (Fig.~\ref{fig:transportsetup}), and in the remainder of our discussion we work with positive fields only, $B,b \geq 0$. All results obtained for the leakage current $I$ are symmetric in the magnetic field, $I(\boldsymbol{B} ) = I(-\boldsymbol{B} )$. Moreover, we define $g_i^{\pm} = g_i^L \pm g_i^R$ for $i \in \lbrace x,y,z \rbrace$ for later convenience. Before delving into detailed magneto-transport investigations, one may look at a few special cases for the out-of-plane leakage current $I(B)$ in which at least one of the quantities from the set $\lbrace B, b, g^-_z, g^-_x \rbrace$ vanishes. By looking at the representation of the Zeeman term in Eq.~\eqref{eq:HZ} in the singlet-triplet basis,
	\begin{align}
	\label{eq:HZinSTbasis}
		\begin{split}
			H_{\text{Z}} = \sum_{\pm} \pm\frac{B g_z^+}{2}  \vert T_{\pm} \rangle \langle T_{\pm} \vert  + \frac{B g_z^- }{2} \left( \vert S \rangle \langle T_0 \vert +\text{h.c.}\right) \\
			+ \frac{b}{2 \sqrt{2}} \sum_{\pm} \left[ g_x^+  \vert T_0 \rangle \langle T_{\pm} \vert \mp   g_x^-  \vert S \rangle \langle T_{\pm} \vert + \text{h.c.} \right],
		\end{split}
	\end{align}
one can directly read off the cases for which the current vanishes, $I =0$. Additionally, one may read off the cases for which the current does not vanish, $I \neq 0$, but is suppressed since second order transitions are necessary to reach the spin singlet. The results are summarized in Table~\ref{tab:currentcases}. As mentioned in Sec.~\ref{sec:modelandmethods}, one can see that $I=0$ if no magnetic field is applied. Moreover, $I=0$ if the g-tensors in the two dots are equal. The current $I$ also vanishes when the applied field only has an out-of-plane component ($b = 0$), but is non-zero when there is only an in-plane component ($B = 0$). The latter case requires second order transitions, and the current is therefore suppressed compared to the case of non-vanishing in-plane and out-of-plane components of the applied magnetic field, which we investigate in the following.

	\begin{table}
		\begin{center}
		\rowcolors{3}{blue!50!white!50}{blue!30!white!40}
 			\begin{tabular}{| m{1.6cm}| m{1.8cm}|| m{1.4cm} | m{2.65cm}|} 
 				\hline
 				\multicolumn{2}{|c||}{$I = 0$}  & \multicolumn{2}{c|}{$I \neq 0$} \\
 				\hline\hline 
				\small{vanishing quantities}   & \small{blocked transition}  &  \small{vanishing quantity} &  \small{necessary second \newline order transition}  \\ 
				 \hline \hline
 				$g^-_z$, $g^-_x $ & $\vert T_{0, \pm} \rangle \rightarrow \vert S \rangle$ & $g^-_z$ & $\vert T_{0} \rangle \rightarrow \vert T_{\pm} \rangle \rightarrow \vert S \rangle$ \\
 				$B$, $b$ & $\vert T_{0, \pm} \rangle \rightarrow \vert S \rangle$ & $B$& $\vert T_{0} \rangle \rightarrow \vert T_{\pm} \rangle \rightarrow \vert S \rangle$ \\
  				$b$ & $\vert T_{\pm} \rangle \rightarrow \vert S \rangle$ & $ g^-_x$ & $\vert T_{\pm} \rangle \rightarrow \vert T_0 \rangle  \rightarrow \vert S \rangle$ \\ 
				 \hline
			\end{tabular}
		\end{center}
	\caption{Value of the out-of-plane leakage current $I(B)$ for given vanishing quantities in the absence of HFI and SOI. The left half of the table shows the cases for which the current $I$ is zero because given quantities vanish (first column) and the corresponding blocked transitions (second column). The right half of the table shows the cases for which the current $I$ is non-zero even though a given quantity vanishes (third column) and the corresponding second order transitions necessary to still reach the singlet (fourth column).  We use the notation $g_z^{\pm} = g_z^L \pm g_z^R$ and $g_x^{\pm} = g_x^L \pm g_x^R$.}
	\label{tab:currentcases}
	\end{table}

\subsection{Out-of-plane magneto-transport}\label{sec:Idifferentgout}
\begin{figure}
		\includegraphics[width=1.0\linewidth]{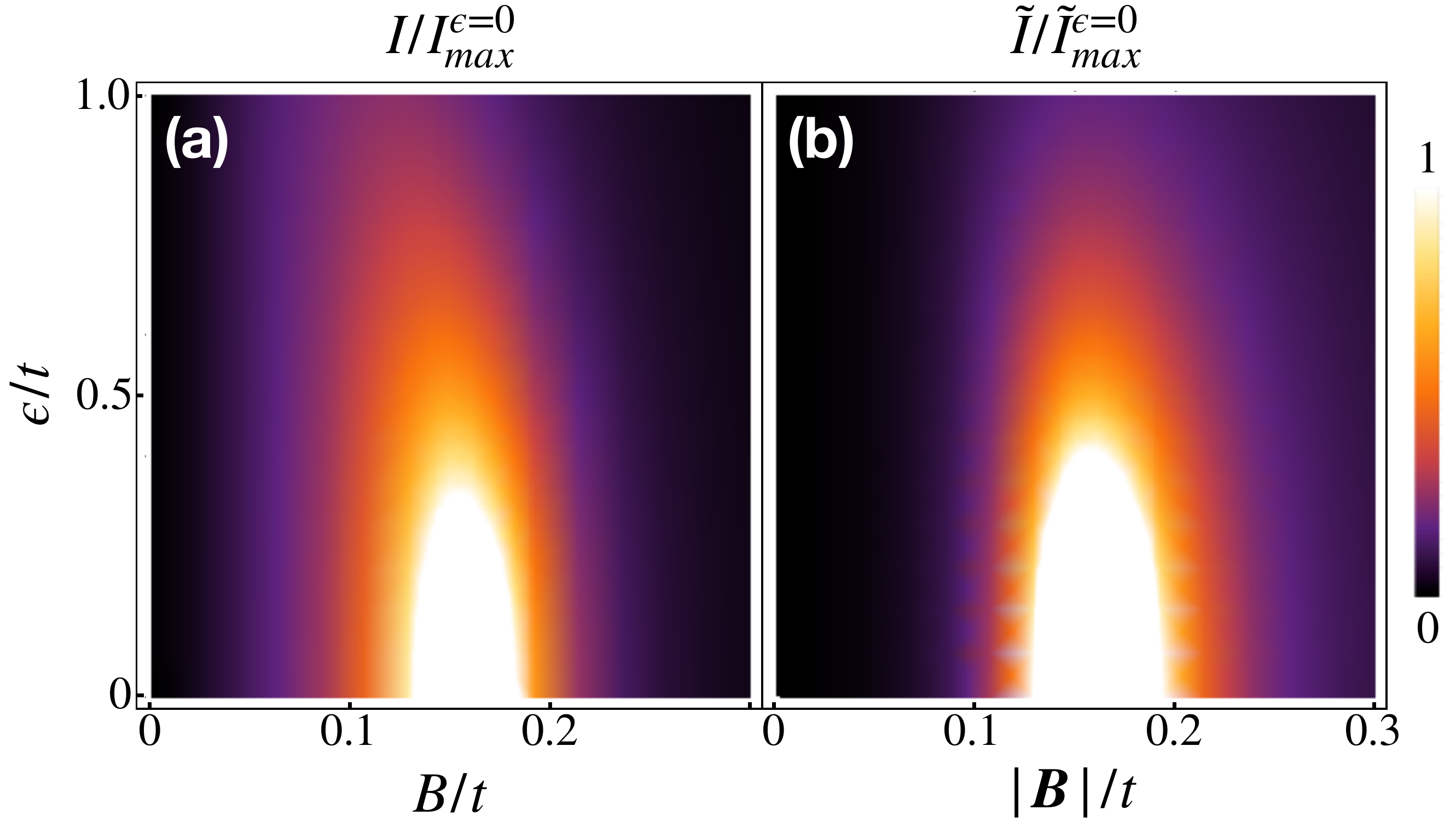}
	\caption{Form of the leakage current. (a) [(b)] Density plot of the leakage current $I$ [$\tilde{I}$] in the absence of HFI and SOI as a function of the detuning $\epsilon$ and the out-of-plane magnetic field $B$ [the magnitude of the magnetic field $ \vert \boldsymbol{B} \vert$] as given by Eq.~\eqref{eq:detuningcurrentperp}. There is a maximum of $I$ [$\tilde{I}$]  as a function of $B$ [$ \vert \boldsymbol{B} \vert$] which is more pronounced at small detunings $\epsilon$. For both plots we use the parameter values $g_z^L = 6.5$, $g_z^R = 6.4$, $g^L_x =0.4$, $g_x^R = 0.3$. For (a) we set $b = 0.1t$ and for (b) we choose the angle $\vartheta = 10^{\circ}$. Performing a fit to the numerical data at $\Gamma_{\text{rel}} =  0$, we obtain (a) $n = 1/4$ and (b) $n = 1/3$. For these values one has (a) $I^{\epsilon = 0}_{\text{max}} = 2.5 \cdot 10^{-4} e \Gamma_R$ and (b) $\tilde{I}^{\epsilon = 0}_{\text{max}} =  1.8 \cdot 10^{-4} e \Gamma_R$.}
	\label{fig:gBanddetuning}
	\end{figure}

For arbitrary detuning $\epsilon$ one may obtain an analytical expression for the leakage current in the regime where the effective in-plane fields and the difference in the out-of-plane fields are much smaller than the energy scale set by the inter-dot tunneling matrix element, i.e., when $\text{max}\left\lbrace g^+_x   b ,   g^-_z  B \right\rbrace \ll t$. In this limit the dominant Hamiltonian is diagonal in the basis
 	\begin{align}
		\left\lbrace \vert T_0 \rangle, \vert T_{\pm} \rangle, \vert S_{\pm} \rangle = \frac{t\vert S \rangle + E_{\pm} \vert S_{02} \rangle} {\sqrt{t^2 + E_{\pm}^2}} \right\rbrace,
	\end{align}
where $E_{\pm} = \left(\epsilon \pm \sqrt{\epsilon^2 +4t^2} \right)/2$ are the energies of the mixed singlets $\vert S_{\pm} \rangle$. The energies of the triplets are zero for $\vert T_0 \rangle$ and $\pm B g_z^+/2$ for the polarized states $\vert T_{\pm}\rangle$.  The system is mostly in one of the triplet states as they are the blocked states at zero field, and the rates of these states decaying via the hybridized singlets into the (0,2) charge configuration may be computed using Eq.~\eqref{eq:rates}.  Combining theses rates with Eq.~\eqref{eq:current} and assuming equal average occupations per charge cycle $n$, one obtains the current
	\begin{equation}
	\label{eq:detuningcurrentperp}
		\frac{I}{e \Gamma_R} = \frac{1}{n} \left[ \left( \frac{2 t}{B g_z^- }\right)^2 + \left( \frac{4t}{b g_x^-} \right)^2 \Lambda(B, \epsilon) \right]^{-1},
	\end{equation}
where
	\begin{align}
		\Lambda(B, \epsilon) =  \left( \left(\frac{B g_z^+ }{2t} \right)^2 - 1 \right)^2 + \left( \frac{\epsilon B g^+_z }{2 t^2} \right)^2.
	\end{align}
The analytical expression \eqref{eq:detuningcurrentperp} can be fitted to the numerical data using the fitting parameter $n$. While $n$ does not depend on the detuning, it does depend crucially on the g-tensors and the in-plane magnetic field and is therefore not a universal quantity. A measurable quantity that does not depend on $n$ is the maximum of the magneto-transport curve $I(B)$. A density plot of the current $I$ as a function of the out-of-plane magnetic field $B$ and the detuning $\epsilon$ is shown in Fig.~\figref[a]{gBanddetuning}. We point out that instead of applying two independent perpendicular fields, one may apply a field of magnitude $B$ at an angle $\vartheta$ to the out-of-plane axis, $\boldsymbol{B} = \vert \boldsymbol{B} \vert (\sin \vartheta, 0 , \cos \vartheta )^T$. Consequently, the current given by 
Eq.~\eqref{eq:detuningcurrentperp} is modified by the the substitutions $B\rightarrow \vert \boldsymbol{B} \vert \cos \vartheta $ and $b \rightarrow \vert \boldsymbol{B} \vert \sin \vartheta $. As can be seen from Fig.~\figref[b]{gBanddetuning}, the altered current is shifted to slightly higher fields. This occurs because the perturbation, which is now a function of the magnitude of the magnetic field $\vert \boldsymbol{B} \vert$ and which couples the polarized triplets to the singlet, increases as $\vert \boldsymbol{B} \vert$ is increased. The existence of a non-zero current in both cases displayed in 
Fig.~\ref{fig:gBanddetuning} shows that we have described a novel channel through which the PSB in a DQD is lifted. Moreover, one can see that $I(B)$ possesses distinct maxima for any given detuning $\epsilon$. In the remainder of this paper, we focus on the case of zero detuning ($\epsilon = 0$) as it shows the most pronounced maximum and deduce information about the g-tensors from the position of this peak. In doing so, we also incorporate the effects of the HFI and SOI on the leakage current.

At zero detuning, $\epsilon = 0$, and under the less restrictive condition $g_x^+   b  \ll t$ one may also obtain an analytical expression for the leakage current. The system's unperturbed eigenstates are the polarized triplets $\vert T_{\pm}  \rangle$ with energies $\pm B g_z^+/2$ and the hybridized states
	\begin{align}
	\label{hybridization}
		\begin{split}
			&\vert \alpha_0 \rangle = \frac{1}{2 E_{\alpha}} \big(B g^-_z \vert  S_{02} \rangle - 2 t \vert T_0 \rangle   \big),\\
			&\vert \alpha_{\pm} \rangle = \frac{1}{\sqrt{8} E_{\alpha}}  \big(2 t \vert  S_{02} \rangle \pm 2E_{\alpha} \vert S \rangle + B g^-_z \vert  T_0 \rangle \big),
		\end{split}
	\end{align}
with energies zero and $\pm E_{\alpha} \equiv \pm \sqrt{(B g^-_z/2)^2 + t^2}$, respectively. The rates of the decoupled triplet states transitioning into the (0,2) singlet via the hybridized states and the rates of the hybridized states reducing to the (0,2) singlet again may be computed from Eq.~\eqref{eq:rates}, and the current $I(B)$ is obtained from Eq.~\eqref{eq:current}. The resulting expression, which is shown in Appendix~\ref{appx:analyticalcurrent}, possesses a maximum at
	\begin{align}\label{eq:maxcond}
	B^* = \frac{t} {\sqrt{g_z^L g_z^R}}.
	\end{align}
Hence, from a measurement of $B^*$ one can directly relate the out-of-plane g-factors in the dots, and we refer to $B^*$ as the g-tensor resonance of the magneto-transport curve. Physically, the resonance value can be readily understood by considering the energies of the states. While the hybridized state $\vert \alpha_0 \rangle$ has constant energy, the other states' energies change when a magnetic field is applied. At $B =  B^*$ the energies of the polarized triplet states $\vert T_{\pm} \rangle$ are aligned with those of the hybridized states $\vert \alpha_{\pm} \rangle$ (Fig.~\figref[a]{variousplots}). The triplet states, which are blocked at zero magnetic field, can thus resonantly transition to the (0,2) singlet via the hybridized states, and we observe a maximal leakage current. The existence of such magnetic degeneracy points \cite{Palyi2019} is due to the fact that the dominant Hamiltonian in the limit under consideration commutes with the $z$-component of the spin, $S_z = S_z^L + S_z^R$, and therefore the subspaces $S_z = 0$ and $S_z = \pm 1$ do not mix. For small fields, $B \ll t$, the ground state energy is $-t$ in the state $\vert \alpha_- \rangle = (\vert S_{02} \rangle - \vert S \rangle)/\sqrt{2}$, which lies in the $S_z = 0$ subspace. At high fields, $B \gg t$, on the other hand, the ground state is $\vert T_- \rangle$ with energy $-(g_z^L + g_z^R)B$ (for $g_z^{L,R} > 0$), which lies in the $S_z = -1$ subspace. Since the states $\vert \alpha_0 \rangle$ and $\vert T_- \rangle$ cannot mix, the ground state energies must become degenerate at some magnetic field value $B = B^*$.
	\begin{figure}
		\includegraphics[width=1.0\linewidth]{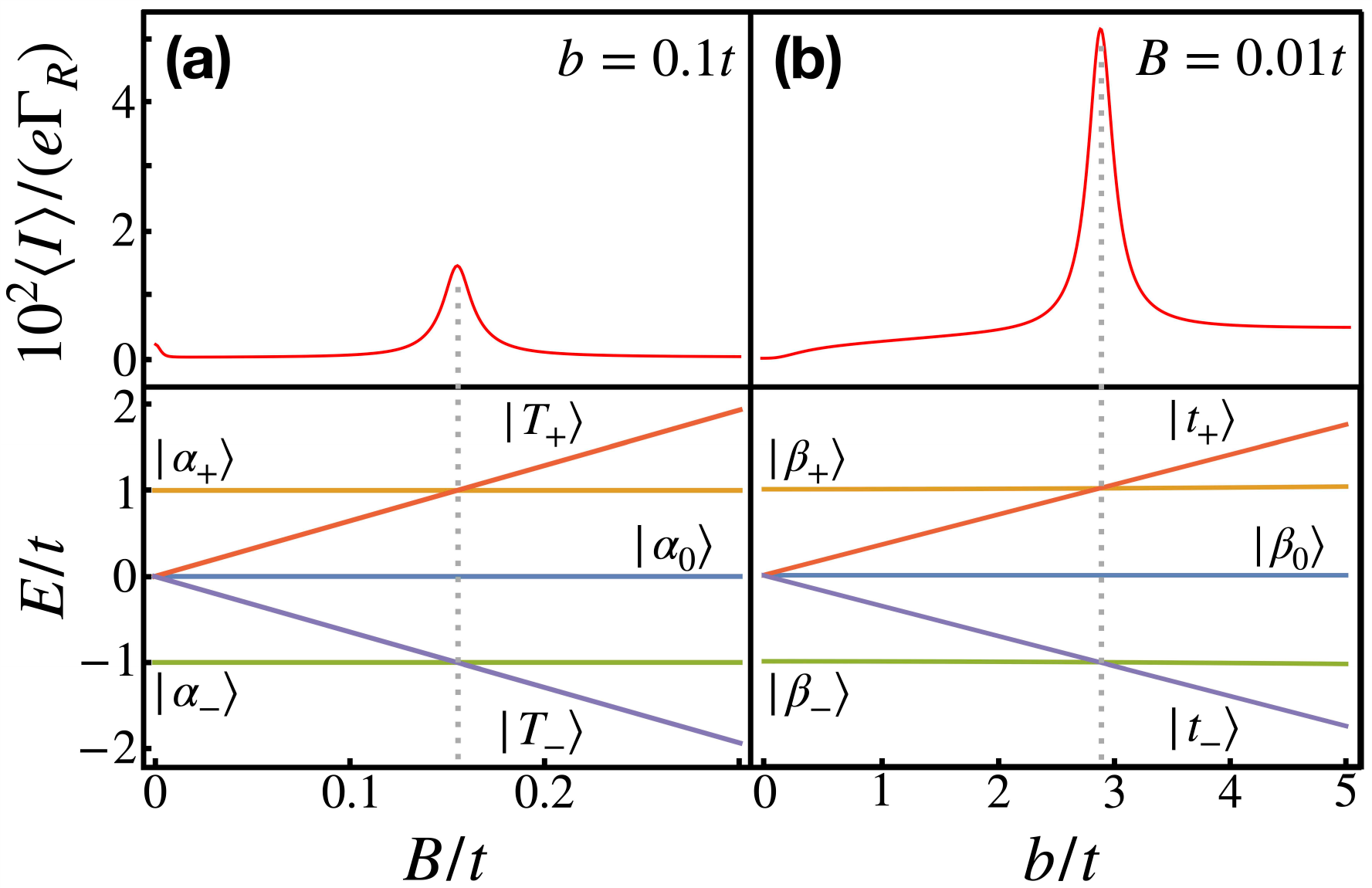}
	\caption{g-tensor resonance of the magneto-transport curve. We show the zero detuning ($\epsilon = 0$) averaged numerical magneto-transport curve $\langle I \rangle$ on top of an energy diagram of the corresponding energy eigenstates of the dominant Hamiltonian  as a function of the (a) out-of-plane and (b) in-plane magnetic field. The g-tensor resonance, which occurs when the triplets can resonantly transition into the singlet-triplet hybridized states, is indicated by a dotted line in each case, and its position can be accurately described with (a) Eq.~\eqref{eq:maxcond} 
	and (b) Eq.~\eqref{eq:inplanemax}. The numerical data was averaged over 100 realizations of the nuclear spins. All plots were created using the same values for the relevant rates and g-tensors, specifically $\Gamma_R =0.1t$, $\Gamma_{\text{rel}} =  0.001t$, $g_x^L = 0.4$, $g_x^R= 0.3$, $g_z^L =6.5$ and $g_z^R =6.4$. To reach the limit of validity of Eq.~\eqref{eq:condHFISOI} we set $\sigma  = 0.001t$ for the standard deviation of the nuclear magnetic fields and $t_{\text{SO}} =0.01t$ for the magnitude of the spin-orbit vector. The applied complementary fields are indicated in the corresponding plot in each case.}
	\label{fig:variousplots}
	\end{figure}

When the system is subject to the HFI and SOI, we must add the terms in Eqs.~\eqref{eq:Hhfi} and \eqref{eq:HSO} to the Hamiltonian, respectively. For the nuclear fields we assume a normal distribution with zero mean as the nuclei are assumed to be unpolarized. The variance of the nuclear spin distribution, $\sigma^2$, is assumed to be equal in both dots.  In Appendix~\ref{appx:maximumcurrent} we show that the maximum of the averaged current occurs to good approximation at the same magnetic field strength  as in the case of vanishing HFI if $(\sigma /B^*)^2 \ll 1$. If true, the interaction with the nuclear spins only causes a broadening of the peak but no shift. The SOI may also move the maximum position away from the analytical prediction for vanishing SOI. To disregard the effects, we must reach the regime $t_{\text{SO}} \ll g_z^+ B^*= t g_z^+/\sqrt{g_z^L g_z^R} $. Since the difference of g-factors is observed to be in the $1 \%$ range \cite{hendrickx2019fast}, $g_z^+ /\sqrt{g_z^L g_z^R}$ is a number of order one, and we require $t_{\text{SO}} \ll t$. To summarize, the conditions which must be fulfilled such that the effects of the HFI and SOI do not change the position of the maximum of the magneto-transport curve can be combined to read
	\begin{align}
	\label{eq:condHFISOI}
		\frac{t_{\text{SO}}}{t}  \ll 1 \ll \frac{ t^2 N_{\text{eff}}}{A^2 \langle I_N \rangle^2}.
	\end{align}
For the sake of the estimate we take $A^2$ now and in the following to be the largest component of the sum $\sum_n A^2_{n,ij}$ and $\langle I_N \rangle$ to be the average nuclear spin of a nucleus in the dots.
\subsection{In-plane magneto-transport}\label{sec:differentgin}
To obtain an analytical expression for the current as a function of the in-plane magnetic field $b$, we work in the regime $ g_z^+ B \ll t$. Since often the out-of-plane g-factors in quantum dots are usually much larger than their in-plane counterparts, e.g., for holes in Ge \cite{Watzinger2016, Winkler2003, Yu2005}, this condition is stronger than the opposite limit discussed in the previous section. If it can be achieved and the inter-dot tunnel coupling is such that the conditions in Eq.~\eqref{eq:condHFISOI} hold, we can obtain an analytical expression for the leakage current as a function of the in-plane magnetic field in complete analogy to the out-of-plane case. This time, the dominant term in the Hamiltonian is diagonal in three states mixing the (1,1) and (0,2) charge configurations directly, $\vert \beta_{0, \pm} \rangle$, with energies $E_0 = 0$ and $E_{S \pm} = \pm \sqrt{(b g^-_x/2)^2 + t^2}$ and two states mixing all (1,1) triplets, $\vert t_{\pm} \rangle$, with energies $E_{t \pm} = \pm b g_x^+/2$. The peak is predicted to be located at
	\begin{align}
	\label{eq:inplanemax}
		b^* = \frac{t}{ \sqrt{g_x^L g_x^R}}.
	\end{align}
This is the value of the field for which two pairs of states (namely $\vert \beta_+ \rangle$ and $\vert t_+ \rangle$ as well as $\vert \beta_- \rangle$ and $\vert t_- \rangle$) are aligned in energy, thus allowing for resonant tunneling of the mixed triplet states into the (0,2) charge state via the states $\vert \beta_{0, \pm} \rangle$ (Fig.~\figref[b]{variousplots}). Measuring the in-plane g-tensor resonance consequently allows us to relate the in-plane g-factors of the dots.
\subsection{The g-tensor resonance in real systems}\label{sec:realsystems}
The conditions in Eq.~\eqref{eq:condHFISOI} contain information about the range in which the effects of site-dependent g-tensors on the leakage current dominate over the effects of the SOI and HFI in a region around the g-tensor resonance. To see whether these conditions can be achieved in present day experiments, we study the cases of (i) electrons in GaAs and (ii) holes in Ge DQDs.

(i) For electrons in GaAs one has $l_{\text{SO}} \approx 3.5$ $\mu$m \cite{Nichol2015}. The estimate $t_{\text{SO}}/t \approx r_{\text{dot}}/l_{\text{SO}}$, where $ r_{\text{dot}}$ is the typical dot radius (of the order of a few $10$ nm) and $ l_{\text{SO}}$ is the particle's spin-orbit length, yields $ t_{\text{SO}}/t \approx 0.01$. To estimate the HFI, one may set $N_{\text{eff}} = 10^5$ for the effective number of nuclei in each dot, $\langle I_N \rangle \approx 3/2$ for the average nuclear spin and $A^2_{\text{GaAs}} \approx 1.2 \cdot 10^{-3}$ (meV)$^2$ \cite{Merkulov2002}. We assume the typical tunnel matrix element $t$ in the DQD to be of the order of $10$ $\mu$eV \cite{Hanson2007} and obtain $ t^2 N_{\text{eff}} /A^2  \langle I_N \rangle^2 \approx 4 \cdot 10^3$. Thus, both conditions in Eq.~\eqref{eq:condHFISOI} from 
Sec.~\ref{sec:Idifferentg} are satisfied for standard quantum dots, making it possible to measure the g-tensor resonance in GaAs DQDs with present day technology. We point out that even though we consider a system with relatively strong HFI (e.g. compared to hole systems), the nuclear magnetic fields do not perturb the g-tensor resonance.

(ii) Recent experiments suggest an out-of-plane g-factor $g_z \approx 6.5$ and an in-plane g-factor $g_x \approx 0.4$ for heavy holes in Ge \cite{Watzinger2018, hofmann2019assessing, Hendrickx2020}.  To estimate the HFI, we use the values $N_{\text{eff}} = 10^5$ \cite{burkardreview2}, $\langle I_N \rangle \approx 9/2 \cdot 7.8 \% = 0.35$ and $A^2_{\text{Ge}} \approx 0.025$ ($\mu$eV)$^2$ \cite{Koh1985}. This yields $t^2 N_{\text{eff}} /A^2  \langle I_N \rangle^2 \approx 10^9 \gg 1$ as required by Eq.~\eqref{eq:condHFISOI} for the HFI. Consequently, the standard deviation of the nuclear magnetic fields can be estimated as $\sigma = 0.001t$ (Fig.~\ref{fig:variousplots}). Concerning the SOI, we note that for holes in Ge hut wires one has the bound  $l_{\text{SO}} \leq 20$ nm \cite{Higginbotham2014}, yielding $ t_{\text{SO}}/t \lesssim 1$ even for small dots with radius $r \lesssim 20$ nm. Exact values for other hole systems in Ge are scarce, but we assume the first condition in Eq.~\eqref{eq:condHFISOI} to be violated due to the strong SOI of holes in Ge. Therefore, it is necessary to incorporate the effects of the SOI on the leakage current and the analytical expression for the g-tensor resonance.

\section{Strong spin-orbit interaction}
\label{sec:strongSOI}
In many systems with strong SOI such as a hole DQD in Ge, the first condition in Eq.~\eqref{eq:condHFISOI} is not met 
(Sec.~\ref{sec:realsystems}), and the position of the maximum of the leakage current as a function of the out-of-plane (in-plane) magnetic field is shifted and thus not well described by Eq.~\eqref{eq:maxcond} (Eq.~\eqref{eq:inplanemax}). In this section we consider arbitrary spin-orbit tunneling matrix element vectors $\boldsymbol{t}_{\text{SO}} = (t_x,t_y,t_z)^T$ and show that the simple expressions for the maximum position of the magneto-transport curve in Sec.~\ref{sec:Idifferentg} can be extended to contain the effects of the SOI. For this purpose, first note that the difference of the g-tensor components in the two dots is assumed to be much smaller than their sum, and so the effective magnetic fields are almost parallel. It is therefore helpful to rewrite the Zeeman Hamiltonian in Eq.~\eqref{eq:HZ} in terms of the sums and differences of the effective magnetic fields in the left ($L$) and right ($R$) dot,
	\begin{align}
		\begin{split}
			H_{\text{Z}} = \frac{1}{2} \left( \boldsymbol{\mathcal{B}}^L + 	\boldsymbol{\mathcal{B}}^R \right) \left( \boldsymbol{S}^L + \boldsymbol{S}^R \right) \\
			+ \frac{1}{2} \left( \boldsymbol{\mathcal{B}}^L - \boldsymbol{\mathcal{B}}^R \right) \left( \boldsymbol{S}^L - \boldsymbol{S}^R \right).
		\end{split}
	\end{align}
In the presence of SOI and HFI the Hamiltonian of the system at zero detuning, $\epsilon = 0$, reads
	\begin{align}
		\begin{split}
			H =  i \boldsymbol{t}_{\text{SO}}  \cdot \vert \boldsymbol{T} \rangle \langle S_{02} \vert + t \vert S \rangle \langle S_{02} \vert + H_{\text{Z}} + H_{\text{HFI}},
		\end{split}
	\end{align}
where
	\begin{align}
		\vert \boldsymbol{T} \rangle = \begin{pmatrix}
  \vert T_x \rangle \\   
  		\vert T_y \rangle \\   
  		\vert T_z \rangle \end{pmatrix} = \begin{pmatrix} \left(\vert T_- \rangle - \vert T_+ \rangle \right)/\sqrt{2}  \\
  		i \left(\vert T_- \rangle + \vert T_+ \rangle \right)/\sqrt{2} \\ 
  		\vert T_0 \rangle \end{pmatrix}.
	\end{align}
We now apply a magnetic field of strength $B$, polar angle $\alpha$ and azimuthal angle $\varphi$ (as measured in a coordinate system in which the out-of-plane axis is defined as the zenith) such that $\boldsymbol{\mathcal{B}}^L + \boldsymbol{\mathcal{B}}^R$ is parallel to the spin-orbit vector $\boldsymbol{t}_{\text{SO}}$. A schematic DQD setup displaying one possible such configuration is shown in Fig.~\figref[a]{SOIcurrent}. The direction of the magnetic field is related to the orientation of  $\boldsymbol{t}_{\text{SO}}$ via 
	\begin{subequations}
	\begin{align}
		&\tan \alpha =   \frac{\sqrt{(t_x/g_x^+)^2 + (t_y/g_y^+)^2}}{t_z/g_z^+} , \\
	 	& \tan \varphi = \frac{t_y/g_y^+}{t_x/g_x^+} .
	\end{align}
	\end{subequations}
Next, we decompose $\boldsymbol{\mathcal{B}}^L - \boldsymbol{\mathcal{B}}^R$ into its components parallel and perpendicular to $\boldsymbol{t}_{\text{SO}}$ and treat the latter as a perturbation 
(Appendix~\ref{appx:validitySOI}).  Finally, choosing the quantization axis to be parallel to the spin-orbit vector $\boldsymbol{t}_{\text{SO}}$ and denoting this direction by the unit vector $\hat{\boldsymbol{n}}$, the dominant Hamiltonian, $H_0$, takes the form
	\begin{align}
	\label{eq:H0strongSOI}
		\begin{split}
			& H_0 =   i t_{\text{SO}}  \vert T_{\hat{\boldsymbol{n}}} \rangle \langle S_{02} \vert + t \vert S \rangle \langle S_{02} \vert \\
			& + \frac{g^+_{\alpha, \varphi}B}{2}  \left(S^L_{\hat{\boldsymbol{n}}} +S^R_{\hat{\boldsymbol{n}}⁄}  \right) + \frac{g^-_{\alpha, \varphi}B}{2} \left(S^L_{\hat{\boldsymbol{n}}} - S^R_{\hat{\boldsymbol{n}}} \right),
		\end{split}
	\end{align}
where $t_{\text{SO}} = \vert \boldsymbol{t_{\text{SO}}} \vert$ and
	\begin{subequations}
		\begin{align}
			\label{eq:galpha+}
	 		& g_{\alpha, \varphi}^+ = \sqrt{\left( g_{\parallel, \varphi}^+ \sin \alpha  \right)^2 + \left( g_{z}^+ \cos \alpha \right)^2}, \\
	 		& g_{\parallel, \varphi}^+ = \sqrt{\left( g^+_x \cos  \varphi  \right)^2 + \left( g^+_y \sin \varphi \right)^2}, \\
	 		&g_{\alpha, \varphi}^- =\frac{g^{+-}_{\parallel, \varphi} \sin^2   \alpha + g_z^+ g_z^- \cos^2 \alpha  }{g_{\alpha, \varphi}^+}, \\
	 		\label{eq:gparallel+-}
	 		& g^{+-}_{\parallel, \varphi} = g_x^+ g_x^- \cos^2   \varphi + g_y^+ g_y^- \sin^2  \varphi .
		\end{align}
	\end{subequations}
The energy spectrum of the Hamiltonian in Eq.~\eqref{eq:H0strongSOI} can be obtained exactly. Considering our findings in Secs.~\ref{sec:Idifferentgout} and \ref{sec:differentgin}, we expect the g-tensor resonance to occur when two pairs of states - each consisting of a state mixing the triplet and singlet subspaces and a mixed triplet state - align in energy. This condition is satisfied when the magnitude of the applied magnetic field takes the resonant value
	\begin{align}
	\label{eq:maxcondwithSOI}
		B^* = \sqrt{\frac{t^2+ t_{\text{SO}}^2}{\left( g_{\alpha, \varphi}^+/2 \right)^2 - \left( g_{\alpha, \varphi}^-/2 \right)^2}}.
	\end{align} 
Note that when the spin-orbit vector $\boldsymbol{t}_{\text{SO}}$ is not aligned with one of the g-tensors' principal axes, there is a finite perpendicular effective magnetic field which we can treat as a perturbation. When $\boldsymbol{t}_{\text{SO}}$ is along one of principal axes or the g-tensors in the dots are isotropic, the component of $\boldsymbol{\mathcal{B}}^L - \boldsymbol{\mathcal{B}}^R$ perpendicular to $\boldsymbol{t}_{\text{SO}}$ vanishes. As a consequence, one must apply a second external field perpendicular to the original one to obtain a non-zero current since the SOI only produces transitions between the singlet and the triplet that is already coupled due to the effects of different g-tensors. The remaining transitions from the triplet subspace to the singlet in the (1,1) charge configuration must be induced by a site-dependent effective magnetic field perpendicular to the original field (see Table~\ref{tab:currentcases}). Clearly, the direction of the complementary effective field is the same as the applied field in both dots in these cases. We recognize the special cases of a dominant magnetic field along one of the g-tensors' principal axes as corresponding to the out-of-plane 
(Sec.~\ref{sec:Idifferentgout}, $\alpha = 0$) and in-plane (Sec.~\ref{sec:differentgin}, $\alpha = \pi/2$) magneto-transport investigations discussed previously. Indeed, the denominator in 
Eq.~\eqref{eq:maxcondwithSOI} reduces to the denominator in Eq.~\eqref{eq:maxcond} 
(Eq.~\eqref{eq:inplanemax}) for $\alpha = 0$ ($\alpha = \pi/2$).

\begin{figure}[t]
		\includegraphics[width=0.97\linewidth]{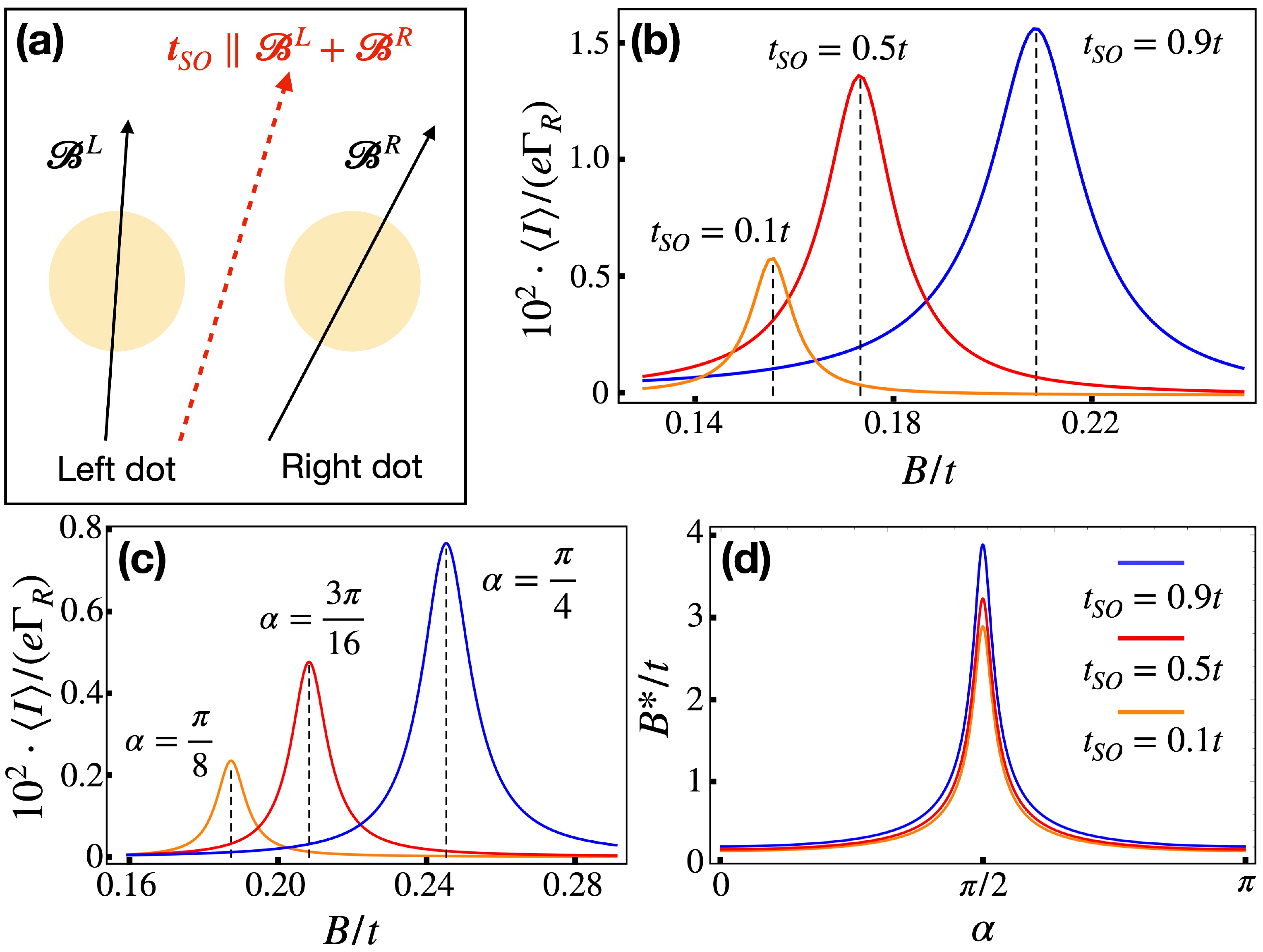}
	\caption{Magneto-transport in the presence of strong SOI. (a) Schematic of a DQD system with site-dependent effective magnetic fields $\boldsymbol{\mathcal{B}}^{L,R}$ and spin-orbit induced inter-dot tunneling. We apply a magnetic field such that $\boldsymbol{\mathcal{B}}^L + \boldsymbol{\mathcal{B}}^R$ is parallel to the spin-orbit vector $\boldsymbol{t}_{\text{SO}}$. (b),(c) The averaged numerical leakage current $\langle I \rangle$ at zero detuning, $\epsilon = 0$, as a function of the magnitude of the magnetic field, $B$, for (b) different strengths of the SOI and $\boldsymbol{t}_{\text{SO}} = (0,0,t_{\text{SO}})^T$ and (c) different values of the polar angle $\alpha$ and $t_{\text{SO}} = 0.5t$. Since the spin-orbit vector is parallel to one of the g-tensors' principal axes in (b), we need to apply a separate perpendicular field, $b = 0.1t$, while this is not necessary in (c). All curves were obtained by averaging over 100 realizations of the nuclear spins, which are described by a normal distribution with zero mean and standard deviation $\sigma = 0.001t$. The dashed lines in (b), (c) indicate the maximum position as determined from the analytical expression in 
	Eq.~\eqref{eq:maxcondwithSOI}. (d) The position of the maximum of the magneto-transport curve as a function of the polar angle $\alpha$ for different magnitudes of the spin-orbit vector, $t_{\text{SO}}$. The parameters used to create all curves in (b) - (d) are $\Gamma_{\text{rel}} = 0.001t$, $\Gamma_R = 0.1t$, $g_z^L = 6.5$, $g_z^R = 6.4$, $g_x^L = 0.4$ and $g_x^R = 0.3$.}
	\label{fig:SOIcurrent}
	\end{figure}

In Fig.~\ref{fig:SOIcurrent} we show the averaged magneto-transport curve as a function of the magnetic field strength $B$ for different values of the magnitude (Fig.~\figref[b]{SOIcurrent}) and different orientations of the spin-orbit vector $\boldsymbol{t}_{\text{SO}}$ (Fig.~\figref[c]{SOIcurrent}). The value of the maximum current increases as the strength of the SOI is grows (Fig.~\figref[b]{SOIcurrent}) since the rate with which triplets can transition to the singlet in the (0,2) charge configuration is increased. Additionally, the value of the maximum current depends on the orientation of the spin-orbit vector (Fig.~\figref[c]{SOIcurrent}) since the perturbing difference in effective magnetic fields causing two of the three triplet-singlet transitions has an angular dependence (Appendix~\ref{appx:validitySOI}) as does the triplet-singlet mixing term ($\sim g_{\alpha, \varphi}^-$) in the dominant Hamiltonian $H_0$ in Eq.~\eqref{eq:H0strongSOI}. 
Finally, Fig.~\figref[d]{SOIcurrent} shows the maximum position as predicted by Eq.~\eqref{eq:maxcondwithSOI} for the complete range of polar angles $\alpha \in [ 0, \pi ]$. The analytical expression in Eq.~\eqref{eq:maxcondwithSOI} predicts the position of the maximum of the leakage current accurately. This yields three scenarios of how information can be extracted from the g-tensor resonance: (i) One may obtain the magnitude of the spin-orbit vector, given its orientation and the g-factors in the dots are known  (Fig.~\figref[b,d]{SOIcurrent}). The orientation can be obtained from other transport investigations as described in Ref.~\cite{Danon2009}, while the g-tensors can be extracted from the Kondo splitting in magneto-conductance investigations \cite{Csonka2008}, Zeeman energy measurements \cite{Witek2011, Brauns2016} or electron dipole spin resonance lines \cite{Watzinger2018, Hendrickx2020}. (ii) Using the angular dependence of the maximum position, one may obtain the orientation of the spin-orbit vector if its magnitude and the g-tensors in the dots are known (Fig.~\figref[c,d]{SOIcurrent}). (iii) Finally, if the orientation and magnitude of the spin-orbit vector are known, we can gain information about the g-tensors in the dots, thus extending the range of validity of the methods discussed in Sec.~\ref{sec:Idifferentg} to include systems with strong SOI such as hole DQDs in Ge.

\section{Conclusion}\label{sec:conclusion}
In conclusion, we show that the PSB in a DQD system is lifted by site-dependent g-tensors and argue that this effect can be more pronounced than the effects of the HFI and SOI. By deriving analytical expressions for the leakage current in special parameter regimes, we are able to relate the g-tensor components in the two dots to the value of the magnetic field for which the leakage current assumes its maximum. Finally, these results are extended to contain the effects of strong SOI, and we point out the prospect of extracting information about the g-tensors and SOI of realistic quantum systems from magneto-transport measurements.

Hole DQDs in Ge present a particularly promising candidate for measuring the g-tensor resonance if information about the spin-orbit vector is available. Hence, further work in this direction should be devoted to exploring the SOI in specific systems such as Ge nanowires and planar heterostructures. Moreover, with the aim of all-electrical control over spins in DQDs as a long-term goal, future  investigations of the PSB may consider the effects of electric fields on the leakage current via the electric dipole spin resonance mechanism and cavity quantum electrodynamics.

\section{Acknowledgements}
We acknowledge support by the DFG through SFB 767.

\appendix

\section{Analytical expression for the leakage current at zero detuning}\label{appx:analyticalcurrent}
In this appendix, we display the full expression for the out-of-plane leakage current at zero detuning, $\epsilon = 0$, in the regime $ g_x^+   b  \ll t$. As in the main text we  denote $g_z^{\pm} = g_z^L \pm g_z^R$ and $g_x^{\pm} = g_x^L \pm g_x^R$. When a magnetic field of the form $\boldsymbol{B} = (b, 0, B)^T$ is applied, the leakage current assumes the form
	\begin{widetext}
		\begin{align}
			\begin{split}
 				\left( \frac{I}{e \Gamma_R } \right)^{-1} =  n_{\alpha} \bigg( 5 + \frac{4t^2}{B^2 (g_z^-)^2} +\frac{B^2 (g_z^-)^2}{t^2} \bigg) + \left(B^2 g_z^L g_z^R - t^2 \right)^2 \sum_{\pm} n_{\pm}  \frac{2(B ^2 (g_z^-)^2 + 4t^2)^2 g_z^+}{t^2}  \\
 				\times \left[ b^2 \left( \frac{B^2}{2 \sqrt{2}} g_x^+ g_z^- (g_z^+)^2 - \frac{1}{2  \sqrt{2}} g_x^- g_z^+\left(B^2(g_z^-)^2 +4 t^2 \right) \mp g_x^+ g_z^- \left(B^2 g_z^L g_z^R - t^2 \right) \right)^2 \right]^{-1}  ,
			\end{split}
		\end{align}
	\end{widetext}
where we distinguish between the average occupation numbers of the $\vert T_{\pm} \rangle$ states ($n_{\pm}$) and those of the hybridized states $\vert \alpha_i \rangle$ (assumed to be equal and denoted by $n_{\alpha}$). The formula predicts a maximum of the current when $B^2 g_z^L g_z^R - t^2 = 0$. This condition is met when the out-of-plane magnetic field takes the value $B^* =  t/\sqrt{g_z^L g_z^R}$. The result for the in-plane leakage current and the g-tensor resonance in the limit $  g_z^+ B \ll  t$ is obtained analogously.

\section{Maximum of the average current}\label{appx:maximumcurrent}
In this appendix we investigate under which conditions the position of the maximum of the leakage current due to site-dependent g-tensors is unchanged by the HFI. In the main text we average the magneto-transport curve over many realizations of the nuclear spins. In general, one has the condition for the maximum of the average of a multi-variable function $f(x,y)$ over the probability distribution $p(y)$,
	\begin{align}
		\begin{split}
			& \partial_x \langle f(x,y) \rangle_y  = \partial_x \int f(x,y)p(y)dy = \\
			& \int \partial_x  f(x,y)p(y)dy =  \langle  \partial_x f(x,y) \rangle_y  = 0.
		\end{split}
	\end{align}
Expanding $f$ around the distribution's mean $\bar{y}$, one may write $$ 0 = \left\langle \partial_x \sum_{n=0}^{\infty} \frac{c_n(x)}{n!} (y - \bar{y})^n \right\rangle =  \sum_{n=0}^{\infty} \partial_x c_n(x) \frac{\langle (y - \bar{y})^n \rangle }{n!}, $$
where $c_n(x) = \partial_y^n f(x,y) \vert_{\bar{y}}$. Assuming a normal distribution with mean $\bar{y}$ and variance $\sigma^2$, the series simplifies and the condition for the maximum reads
	\begin{align}
	\label{eq:maxcondnew}
		0 = c_0^{\prime} +  \sum_{n \; \text{even}} \frac{c_n^{\prime}}{n!!}  \sigma^n,
	\end{align}
where $c_n^{\prime} = \partial_x c_n$ and we have used the identity $n!! (n-1)!! = n!$. In our analysis of the main text the probability distribution $p \left(K \right)$ of the nuclear magnetic field $K$ is normally distributed with zero mean, and we identify $f(x,y) \rightarrow I \left(B, K \right)$. Moreover, $I$ only depends on the quantity $B + K$ and due to $\bar{K} = 0$ one has $c_0 = I(B, \; \text{HFI} = 0)$ and $c_n^{\prime} = \partial_B^{n+1} I(B, \; \text{HFI} = 0)$. The condition for the maximum of the current $I$ in the absence of HFI is thus $c_0^{\prime} = 0$. Comparing this to the new condition for a maximum, 
Eq.~\eqref{eq:maxcondnew}, we find that the position is in good approximation unchanged if 
	\begin{align}
		\sum_{n \; \text{even}} \frac{c_n^{\prime}}{n!!}  \sigma^n \ll c_0^{\prime}.
	\end{align}
The solution of the stationary master equation in Eq.~\eqref{mastereq} of the main text is a rational function of the magnetic fields. Hence, the terms $c_n^{\prime}(B)$ are of the order $I(B)/B^{n+1}$ and we obtain
	\begin{align} 
		\sum_{n \; \text{even}}  \frac{1}{n!!} \left(\frac{\sigma }{B } \right)^n \ll1.
	\end{align}
Since the above inequality must hold around the maximum position, $B^*$, we find that nuclear spins do not noticeably shift the position of the maximum of the current $I(B)$ if $(\sigma/ B^*)^2 \ll 1$.
\vspace{8mm}

\section{Validity of the perturbative approach to systems with strong SOI}\label{appx:validitySOI}
In this appendix we show that the perturbative approach used in 
Sec.~\ref{sec:strongSOI} of the main text is justified. Working in the setup defined there (in particular using the abbreviations defined in Eqs.~\eqref{eq:galpha+} - \eqref{eq:gparallel+-}), the perturbation term in the Hamiltonian, $H_1$, is given by 
	\begin{align}
		H_1 = \frac{1}{2}\left( \boldsymbol{\mathcal{B}}^L - \boldsymbol{\mathcal{B}}^R \right)_{\perp} \cdot \left( \boldsymbol{S}^L - \boldsymbol{S}^R \right).
	\end{align}
We choose the quantization axes ($\hat{z}^{\prime}$) to be parallel to the sum of effective magnetic fields $\boldsymbol{\mathcal{B}}^L + \boldsymbol{\mathcal{B}}^R$, such that in the new coordinates $(x^{\prime}, y^{\prime}, z^{\prime} )$ the sum of effective magnetic fields becomes $\boldsymbol{\mathcal{B}}^L + \boldsymbol{\mathcal{B}}^R = (0,0, \vert \boldsymbol{\mathcal{B}}^L + \boldsymbol{\mathcal{B}}^R  \vert)^T \equiv (0,0, g_{\alpha, \varphi}^+ B)^T$. The parallel component of the difference in effective magnetic fields reads $\left( \boldsymbol{\mathcal{B}}^L - \boldsymbol{\mathcal{B}}^R \right)_{\parallel} =  (0,0, g_{\alpha, \varphi}^- B)^T$ and the perpendicular component is given by
	\begin{align}
		\left( \boldsymbol{\mathcal{B}}^L - \boldsymbol{\mathcal{B}}^R \right)_{\perp} = \frac{B}{2} \begin{pmatrix}
			\sin 2 \alpha \frac{g_z^- g_{\parallel, \varphi}^+ - g_z^+ g_{\parallel, \varphi}^{+-}/g_{\parallel, \varphi}^+}{g_{\alpha, \varphi}^+} \\
			\sin 2 \varphi \sin \alpha \frac{g_y^+ g_x^- - g_x^+ g_y^-}{g_{\parallel, \varphi}^+} \\
			0
		\end{pmatrix}.
	\end{align}
While the parallel component of $\boldsymbol{\mathcal{B}}^L - \boldsymbol{\mathcal{B}}^R$ is accompanied by the spin operator $S_{z^{\prime}}$, the perturbation appears in the Hamiltonian with the remaining two spin operators $S_{x^{\prime}}$ and $S_{y^{\prime}}$. The strength of the perturbation compared to the dominant magnetic field induced term in $H_0$ is estimated as 
	\begin{align}
	\lambda =	\frac{ \left\vert \left( \boldsymbol{\mathcal{B}}^L - \boldsymbol{\mathcal{B}}^R \right)_{\perp} \right\vert }{ B g_{\alpha, \varphi}^+}.
	\end{align}
For the case of holes in Ge where we assume $g_z^L = 6.5$, $g_z^R = 6.4$, $g_x^L = 0.4$, $g_x^R = 0,3$ and $g_x^{L,R} = g_y^{L,R}$, $\lambda$ has the maximum value $\lambda_{\text{max}} \approx 0.068$ for the polar angle $\alpha = 0.48  \pi$. Thus, $\lambda_{\text{max}} \ll 1$ and the perturbative approach used to obtain the energy spectrum of the DQD system is valid.
\bibliographystyle{apsrev4-2}
\bibliography{PSBliterature2}
\end{document}